\documentclass[twocolumn,trackchanges]{aastex701}

\usepackage[utf8]{inputenc}
\usepackage{float}
\usepackage{comment}
\begin{document}

\title{Gravitationally Lensed View of DSFG-1 in PLCK G165.7+67.0: Strong Dust Emission and Spatially Resolved Stellar Population Analysis with JWST and SMA} 

\author[orcid=0009-0002-7871-3337]{Zhiyu Yan} 
\affiliation{Chinese Academy of Sciences South America Center for Astronomy (CASSACA), National Astronomical Observatories(NAOC), 20A Datun Road, Beijing 100012, China, ydai@nao.cas.cn}
\affiliation{School of Astronomy and Space Science, University of Chinese Academy of Sciences, Beijing 101408, China}
\email{yanzy@bao.ac.cn}

\author[orcid=0000-0001-9773-7479]{Daizhong Liu} 
\affiliation{Purple Mountain Observatory, Chinese Academy of Sciences, 10 Yuanhua Road, Nanjing 210023, China, dzliu@pmo.ac.cn}
\affiliation{State Key Laboratory of Radio Astronomy and Technology, Purple Mountain Observatory, Chinese Academy of Sciences, 10 Yuanhua Road, Nanjing 210023, China}
\email{dzliu@pmo.ac.cn}

\author[orcid=0000-0002-7928-416X]{Y. Sophia Dai} 
\affiliation{Chinese Academy of Sciences South America Center for Astronomy (CASSACA), National Astronomical Observatories(NAOC), 20A Datun Road, Beijing 100012, China, ydai@nao.cas.cn}
%\correspondingauthor{Y. Sophia Dai}
\email{ydai@nao.cas.cn}

\author[0000-0001-9394-6732]{Patrick S. Kamieneski}
\affiliation{Department of Physics and Astronomy, Chalmers University of Technology, SE-412 96 Gothenburg, Sweden}
\email{pkamiene@asu.edu}

\author[0000-0003-2027-8221]{Pierre Cox} 
\affiliation{Sorbonne Université, UPMC Paris 6 and CNRS, UMR 7095, Institut d’Astrophysique de Paris, 98b bd. Arago, 75014 Paris, France}
\email{cox@iap.fr}

\author[0000-0003-1625-8009]{Brenda L. Frye}
\affiliation{Department of Astronomy/Steward Observatory, University of Arizona, 933 N. Cherry Avenue, Tucson, AZ 85721, USA}
\email{bfrye@arizona.edu}

\author[orcid=0000-0003-3032-0948]{QingHua Tan} 
\affiliation{Purple Mountain Observatory, Chinese Academy of Sciences, 10 Yuanhua Road, Nanjing 210023, China}
\email{qhtan@pmo.ac.cn}

\author[orcid=0000-0001-5950-1932]{Fengwei Xu} 
\affiliation{Max Planck Institute for Astronomy, Königstuhl 17, 69117 Heidelberg, Germany}
\email{fengwei@mpia.de}

\author[orcid=0009-0006-4990-7529]{Yixiao Liu} 
\affiliation{Chinese Academy of Sciences South America Center for Astronomy (CASSACA), National Astronomical Observatories(NAOC), 20A Datun Road, Beijing 100012, China, ydai@nao.cas.cn}
\affiliation{School of Astronomy and Space Science, University of Chinese Academy of Sciences, Beijing 101408, China}
\email{liuyixiao@nao.cas.cn}

\author[orcid=0000-0002-7237-3856]{Ke Wang} 
\affiliation{Kavli Institute for Astronomy and Astrophysics, Peking University, 5 Yiheyuan Road, Haidian District, Beijing 100871, China}
\email{kwang.astro@pku.edu.cn}

%\author[orcid=0000-0000-0000-0000]{Collaborators}
%\affiliation{TBD}
%\email{}

\begin{abstract}
We present a detailed stellar population analysis of the strongly lensed dusty star-forming galaxy (DSFG) PLCK~G165.7+67.0 DSFG-1 at $z = 2.236$, combining JWST NIRCam imaging with new Submillimeter Array (SMA) observations.
This source is multiply imaged into two lensed components: image~1a, with a moderate magnification factor of $\mu \sim 5$, and image~1bc, with an extreme magnification factor of $\mu \sim 40$.
The new SMA observations detect significant dust continuum emission at 225\,GHz and 273\,GHz, with combined flux densities of $S_{\rm cont}=(1.19\pm0.38)$~mJy in image~1a and $S_{\rm cont}=(10.02\pm0.85)$~mJy in image~1bc, indicating active star formation at sub-kpc scale.
Based on the integrated SED modeling, DSFG-1 exhibits a lensing amplification-corrected stellar mass of $M_{\star} = (1.2 \pm 0.4) \times 10^{10}~M_{\odot}$, 
and a star-formation rate (SFR) of $(103 \pm 14)~M_{\odot}\,\mathrm{yr^{-1}}$, 
similar to previous $H\alpha$-based results, placing it four times above the star-forming main sequence at this redshift.
Its location on the size–mass plane and its morphological properties suggest that the system occupies a transitional phase between star-forming late-type galaxies and compact early-type systems. Together with its elevated star-formation activity, this is consistent with a rapidly evolving galaxy observed during Cosmic Noon.
We further investigate the spatially resolved stellar population properties, and found significant spatial variations in stellar age and dust attenuation.
These results point to a non-uniform star-formation history and highlight the complex interplay between dust geometry, stellar growth, and gravitational lensing, consistent with a merger scenario.
%These results indicate spatially varying stellar populations, although part of the apparent age variation may be driven by complex dust attenuation, and highlight the interplay between dust geometry, stellar growth, and gravitational lensing, consistent with a merger scenario.
\end{abstract}


\section{Introduction}
\label{sec:introduction}
Dusty star-forming galaxies (DSFGs) at $z \gtrsim 2$ are among the most important contributors to cosmic star formation.
Their intense infrared and submillimeter emission originates from dust-attenuated star formation, but characterizing their internal stellar populations is often hampered by limited angular resolution\citep[e.g.][]{Casey2014}.
Gravitational lensing provides a powerful means of enhancing the effective physical resolution and sensitivity of observations by magnifying background galaxies in both flux and spatial scale resolution.
%Gravitational lensing provides a powerful means of overcoming this limitation by magnifying background galaxies in both flux and spatial scale.
In strongly lensed systems, one can resolve the distribution of dust attenuation, stellar mass, and star formation rate (SFR) across kpc or even sub-kpc scales, offering a unique window into the physical processes governing star formation in the early Universe. \par
High-resolution submillimeter observations over the past decade have significantly revised our understanding of the nature of dusty star-forming galaxies.
Traditionally, submillimeter galaxies (SMGs) and DSFGs have been interpreted as merger-driven starbursts, motivated by their extreme infrared luminosities, compact dust emission, and disturbed gas kinematics \citep[e.g.][]{Tacconi2008}.
In this work, we use the terms SMGs and DSFGs more or less interchangeably, as there is currently no clear consensus on the precise definition of either population.
However, recent high-resolution dust continuum imaging with the Atacama Large sub-millimeter Array (ALMA) has revealed a more nuanced picture.
Studies of unlensed SMGs, most notably the ALMA LABOCA ECDFS Submillimeter Survey (ALESS; \citealt{Hodge2013}) sample, show that while some systems exhibit disturbed morphologies consistent with major mergers, a significant fraction display smooth, centrally concentrated, and disk-like dust emissions, consistent with massive, gas-rich star-forming disks rather than purely merger remnants \citep[e.g.][]{Hodge2016,Hodge2019, Gullburg2019, Gillman2024}. 
%More recently, this picture has been further reinforced by spatially resolved studies that combine ALMA submillimeter imaging with JWST observations.
%Results from the ``[CII] Resolved ISM Star-forming galaxies with ALMA" (CRISTAL) program, together with early JWST/NIRCam imaging, reveal that the dust-continuum morphology traced at submillimeter wavelengths often aligns with extended, structured star-forming components rather than compact merger-driven cores \citep{Herrera-Camus2025}.
Complementary observations combining ALMA and JWST have further demonstrated that the distributions of dust-obscured and unobscured stellar components can differ substantially within individual galaxies, revealing complex internal structures that are not captured by observations at a single wavelength \citep[e.g.][]{Lee2025}. Together, these results suggest that intense dusty star formation at high redshift can arise through multiple physical pathways and underscore the importance of spatially resolved multi-wavelength studies.
Disentangling these scenarios requires spatially resolved measurements of the relative distributions of stars, dust, and star formation, which are generally inaccessible in unlensed systems due to limited angular resolution.
In this context, strongly lensed DSFGs provide a particularly powerful laboratory for probing both the physical origin and spatial distribution of star formation in dust-attenuated galaxies.

PLCK~G165.7+67.0 (hereafter G165) is one such remarkable lensing system, first identified in \textit{Planck} and \textit{Herschel} surveys \citep{Canameras2015, Harrington2016}.
This source is one of the brightest submillimeter objects in the field, a cluster with complex components and structures. 
The submillimeter-bright background source, known as G165 DSFG-1 at $z = 2.236$, is multiply imaged into a northern arc (1a) and a southern, highly stretched arc (1bc) by the foreground cluster at $z = 0.35$ (labeling convention adopted from \citet{Frye2024}).
It is worth noting that unlike 1a, the 1bc arc only incorporates part of the visible galaxy (as shown in \citet{Patrick2024}, Figure~18).  
Previous studies \citep[e.g.][]{Frye2019, Pascale2022, Frye2024, Patrick2024} have revealed that 1bc experiences extreme magnification ($\mu\sim40$), making it one of the brightest known lensed DSFGs. After correcting for lensing amplification, G165 DSFG-1 is inferred to have an infrared luminosity of $L_{\rm IR} \gtrsim 10^{12}L_\odot$, placing it in the class of ultra-luminous infrared galaxies (ULIRGs).\par

%In this work, we use high-angular resolution submillimeter observations with the SMA to investigate the dust morphology and star-forming structure of the extreme starburst PLCK~G165.7+67.0, aiming to assess whether its dust emission is dominated by compact merger-driven components or by more extended disk-like structures.
%We combine the rest-frame near-UV to submillimeter properties of G165 DSFG-1, using high-resolution JWST/NIRCam imaging with new \textit{Submillimeter Array} (SMA) observations and publicly available ALMA data. 
%The SMA data trace the spatially resolved distribution of cold dust and star-forming regions, while the JWST data enable stellar population modeling at comparable resolution. 
%By analyzing both the global SEDs of 1a and 1bc, as well as performing spatially resolved SED fitting through Voronoi binning \citep{Cappellari2003}, we aim to investigate the connection between the stellar and dust components in this lensed DSFG.
In this work, we combine high-resolution JWST/NIRCam imaging with new \textit{Submillimeter Array} (SMA) observations to investigate the relationship between the stellar and dust components of G165 DSFG-1. The SMA observations provide unique spatial information on the warm dust. Combined with the gravitational lensing magnification ($\mu \sim 5$), they enable source-plane reconstruction on sub-kpc scales.
%While the JWST data enable stellar population modeling on comparable physical scales. 
%The SMA imaging further provides an independent consistency check on the adopted lens model through comparisons between the observed and reconstructed continuum emission.
By analyzing both the global SEDs of 1a and 1bc, as well as performing spatially resolved SED fitting through Voronoi binning \citep{Cappellari2003}, we investigate the connection between obscured star formation and stellar populations in this strongly lensed DSFG.

%Throughout this paper, we adopt the source labeling convention introduced by \citet{Frye2024}, in which the two components are referred to as 1a and 1bc.
%The northern component of G165 has been identified as Arc~1a in previous analyses. A compact knot located $\sim0\farcs22$ to the west, labeled NS\_46 in recent JWST catalogs, appears morphologically and photometrically consistent with the 1a arc.
%Considering their continuity and the lensing configuration, we treat both as a single system and refer to them collectively as “1a” throughout this paper.\par
Previous studies distinguished the northern lensed structure into two components: Arc~1a, corresponding to the southern DSFG component, and NS\_46, a spectroscopically confirmed companion located immediately to the north \citep{Frye2024}. In the JWST imaging, however, the two components appear directly connected with no clear morphological boundary. For simplicity in the following analysis, we refer to the entire connected northern structure collectively as “1a”, while retaining the original nomenclature when discussing results from previous studies.\par
This paper is organized as follows: In Section 2, we describe the datasets and observations used in this study.
In Section 3, we present the SED analysis, including the global SEDs of images 1a, 1bc and system DSFG-1, as well as the spatially resolved SEDs.
In Section 4, we provide a discussion of the results.
In Section 5, we summarize our main conclusions of this paper.
Throughout this paper, we adopt a flat $\Lambda$CDM cosmology with $h \equiv H_{0} / (100\,\mathrm{km\,s^{-1}\,Mpc^{-1}}) = 0.7$, $\Omega_{\mathrm{M}} = 0.3$, and $\Omega_{\Lambda} = 0.7$.

\section{Data} \label{sec:data}
\subsection{SMA} 
New Submillimeter Array observations for G165 DSFG-1 were conducted in the Extended configurations on July 12th and July 19th ,2022, with a total on-source exposure time of $\sim7.6$\,hr (2022A-S036, PI: Fengwei Xu).
Time-dependent antenna gains were monitored by periodic observations of the calibrator 1159+292; frequency-dependent bandpass responses were calibrated by the calibrator BLLac; the absolute flux was scaled by observed correlator counts with modelled fluxes of Neptune.
The baseline lengths range from 62 m to 226 m, which corresponds to the maximum recoverable scale (MRS) of $\sim 4\farcs3$ and the expected angular resolution of $\sim1\farcs2$ at 230 GHz.
The full width at half-maximum (FWHM) of the primary beam is about $\sim44\farcs8$ at 230 GHz. \par
For the two tracks, we used the same setup for the upgraded SWARM correlator in the Dual Rx mode, where two receivers Rx240 and Rx345 are used to cover a bandwidth of 12 GHz per sideband per receiver (i.e., 48 GHz in total).
The LO frequencies of Rx240 and Rx345 are 225 GHz and 273 GHz respectively, with a uniform channel width of 139.6 kHz (equivalent velocity 0.18 $km~s^{-1}$ at 230 GHz) across the entire band. System temperatures varied from 100 to 150 K, and the zenith opacity at 225 GHz was around 0.1 during the tracks. Five antenna were employed on July 12th and 6 on July 19th.\par
\begin{figure*}[t]
    \centering    \includegraphics[width=\textwidth]{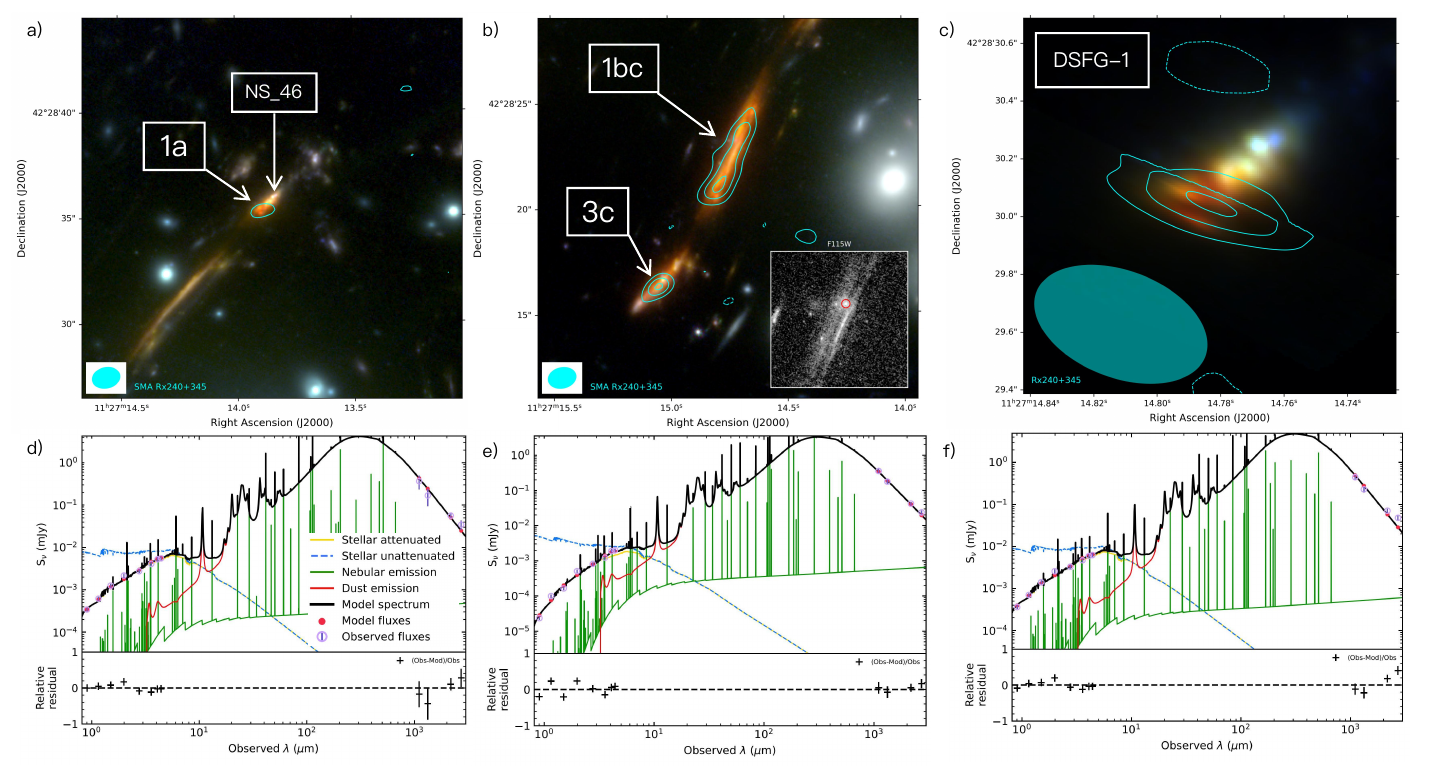}
    \caption{
    JWST NIRCam RGB images (F115W, F277W, and F444W) overlaid with SMA continuum contours for the lensed components of G165-DSFG-1 and the corresponding SED fitting results.
    Panels \textbf{\textit{a)}} and \textbf{\textit{b)}} show the JWST RGB images of the northern Arc~1a and the southern Arc~1bc in the image plane, respectively, with SMA continuum contours overlaid.
    The inset in panel \textbf{\textit{b)}} shows the corresponding JWST F115W grayscale image of Arc~1bc. The red circle marks a likely foreground object projected onto Arc~1b, which is not associated with the lensed source.
    The Arc~3c of DSFG-3 is also visible in \textbf{\textit{b)}} and is labeled accordingly.
    Panels \textbf{\textit{d)}} and \textbf{\textit{e)}} present the best-fit SEDs of arcs 1a and 1bc derived in the source plane.
    Panel \textbf{\textit{c)}} shows the reconstructed source-plane JWST RGB image with SMA continuum contours for the combined DSFG-1 system, while panel \textbf{\textit{f)}} presents the global SED fitting result of DSFG-1.
    In panels \textbf{\textit{a)}} and \textbf{\textit{b)}}, SMA contours are shown at $-4\sigma$ (dashed) and $4\sigma$, $7\sigma$, $10\sigma$ (solid), where $\sigma = 7.892$ MJy\,sr$^{-1}$. In panel \textbf{\textit{c)}}, SMA contours are shown at $-2\sigma$ (dashed) and $4\sigma$, $6\sigma$, $8\sigma$ (solid), with the same noise level $\sigma = 7.892$ MJy\,sr$^{-1}$.
    The source-plane beam, indicated in the lower left corner of panel \textbf{\textit{c)}}, is derived by projecting the image-plane SMA beams of arcs 1a and 1bc through the gravitational lensing model and averaging the resulting source-plane beams. 
    }
    \label{fig:SMA}
\end{figure*}

The raw data are first converted into measurement set, which are available for CASA \citep{McMullin2007, Casa2022} and then calibrated using CASA 6.6.1 manually. To maximize the signal-to-noise ratio, we combine the data in two tracks and in two receivers. We produced three continuum products: (1) a map of Rx240 only, with synthesized beam of $1\farcs39 \times 1\farcs02$ at PA = $-76.39^\circ$, (2) a map of Rx345 only, with synthesized beam of $1\farcs13 \times 0\farcs82$ at PA = $-73.42^\circ$, and (3) a combined map using both receivers, with synthesized beam of $1\farcs30 \times 0\farcs95$ at PA = $-76.74^\circ$. The single-receiver maps serve as independent photometric measurements in the global SED analysis, whereas the combined (Rx240+Rx345) map, which provides the highest signal-to-noise ratio and best imaging fidelity, is adopted for morphological analysis and comparison with JWST imaging. 
In Figure~\ref{fig:SMA}, we show the SMA continuum contours overlaid on the JWST/NIRCam imaging of G165-DSFG-1, with panels (a) and (b) displaying the northern image 1a and the southern highly magnified Arc~1bc, respectively.
The SMA continuum emission is clearly detected toward both components.
Arc~1a is detected at $\sim6\sigma$ level, with a flux density of $(1.19 \pm 0.38)\,$mJy.
Arc~~1bc is detected at $\sim11\sigma$ level, with a flux density of $(10.02 \pm 0.85)\,$mJy.
In Arc~1bc, the SMA continuum emission broadly overlaps with the NIRCam morphology.
In contrast, Arc~1a shows a more pronounced offset between the SMA and NIRCam emission, with a noticeable difference in orientation.
This discrepancy is most likely driven by the elongated synthesized beam and limited angular resolution of the SMA observations, although intrinsic differences between the distributions of cold dust and stellar emission may also contribute.\par
In addition to the continuum emission associated with DSFG-1, a secondary SMA detection is present to the southeast of Arc~1bc at a significance level of $\sim7\sigma$, with a flux density of $(2.60 \pm 0.45)\,$mJy. This source coincides with Arc~3c of the independently lensed system DSFG-3 identified in previous studies. As it is not associated with DSFG-1, we do not consider it further in the analysis presented in this work.

\subsection{JWST NIRCAM} 
We make use of science-ready reduced JWST NIRCam imaging of PLCK G165.7+67.0 from the DAWN JWST Archive (DJA), which provides fully calibrated, science-ready mosaics combining all available exposures from programs PID 1176 (PI: Windhorst) on March 30, 2023 and PID 4446 (PI: Frye) on April 22 and May 9, 2023 respectively.
The DJA products are resampled onto a common pixel grid with a pixel scale of $0.05\arcsec$ per pixel, and photometry is then measured as described in Section~\ref{sec:photometry}.
We use eight NIRCam filters, namely F090W, F115W, F150W, F200W, F277W, F356W, F410M, and F444W, covering the rest-frame near-UV to near-infrared emission at the source redshift ($z=2.236$). 
A three-color JWST/NIRCam composite image, constructed from the F115W, F277W, and F444W bands and overlaid with the combined Rx240+345 SMA continuum contours, is shown in Figure~\ref{fig:SMA} (panel a and b), and the SMA continuum contours, as described in the previous section, are overlaid to trace the dust distribution relative to the stellar populations revealed by JWST.
Both lensed components exhibit complex, clumpy near-infrared morphologies, with multiple local brightness peaks that vary across NIRCam bands.
In the dust continuum submillimeter emission, Arc~1a appears relatively compact and irregular, whereas Arc~1bc is stretched into an elongated arc, with several brightness peaks distributed along the magnification direction.
We defer a detailed physical interpretation of these morphological features to Section~\ref{sec:discussion}.

\subsection{ALMA} 
We also retrieved the publicly available ALMA observations of G165 DSFG-1 from the ALMA Science Archive. The dataset corresponds to project 2021.1.00607.S (PI: Cañameras), with observations carried out in Band~3 ($\lambda \approx 2.7$\,mm; 111\,GHz) and Band~4 ($\lambda \approx 2.2$\,mm; 138\,GHz), which was described and published in \citet{Patrick2024}.
We re-imaged the calibrated data to produce continuum maps. The synthesized beam sizes are $1\farcs65 \times 0\farcs69$ with  PA = $-18.4^\circ$ in Band~4, and $1\farcs98 \times 0\farcs89$ with a position angle of PA = $-2.7^\circ$ in Band~3. \par
Significant dust continuum emission is detected in both bands. In Band~4 (2\,mm), we measure continuum flux densities of $(122 \pm 25)\,\mu$Jy for Arc~1a and $(873 \pm 49)\,\mu$Jy for Arc~1bc, corresponding to detections at $\sim6\sigma$ and $\sim15\sigma$, respectively. In Band~3 (3\,mm), the corresponding flux densities are $(181 \pm 20)\,\mu$Jy (1a) and $(1603 \pm 46)\,\mu$Jy (1bc), corresponding to detections at $\sim12\sigma$ and $\sim28\sigma$, respectively.
Figure~\ref{fig:contour_ALMA} in Appendix~\ref{sect:ALMA} presents the Band~3 and Band~4 continuum contours overlaid on the JWST NIRCam RGB images in both the image plane and the reconstructed source plane.\par
In this work, the ALMA continuum measurements provide constraints on the cold dust emission of G165 DSFG-1 and anchor the long-wavelength side of the spectral energy distribution, complementing the JWST NIRCam near-infrared observations.

\subsection{Lensing Model} 
We adopt the best-fitting gravitational lensing model of the PLCK G165.7+67.0 cluster as presented in \citet{Patrick2024} (see also \citet{Pascale2025}, Appendices A and B, for more details on this model and other models for the same cluster field). The model provides image-to-source-plane mappings and magnification estimates, which are used for source-plane reconstruction and for correcting photometry and physical parameters. 
For all source-plane reconstructions, we adopt an oversampled grid with a pixel scale of $0\farcs00625$.
Figure~\ref{fig:SMA} panel~\textbf{\textit{c)}} shows the source-plane reconstructed JWST NIRCam RGB image composed of F115W (blue), F277W (green), and F444W (red). Overlaid is the combined SMA continuum map (Rx240+345) reconstructed into the source plane.
The source-plane reconstruction is derived from a joint lens-model inversion of all available lensed images. We verified that the source morphology inferred from images 1a and 1bc is consistent with the combined reconstruction shown here.
The approximate source-plane beam $0\farcs64 \times 0\farcs37$ at $\mathrm{PA}=69.37^\circ$ is shown in the lower left corner of the panel.

Magnification varies across filters due to wavelength-dependent morphology and differential lensing. Therefore, for the global SED analysis, photometry was performed directly in the reconstructed source plane rather than applying band-by-band magnification corrections, ensuring the most accurate spectral measurements.
Previous Lenstool-based lens modeling \citep{Patrick2024} found that the effective magnification factors of image~1a are remarkably stable across different bands, with variations of only $\sim2\%$, while image~1bc exhibits somewhat larger variations of up to $\sim10\%$. To further assess the impact of differential lensing on the inferred magnification factors, we performed an independent MCMC-based forward-modeling analysis (see Section~\ref{sec:Re}) of the dust continuum emission. 
%Although the primary goal of this modeling is to constrain the intrinsic source morphology, it also provides posterior distributions for the effective magnification factors, allowing an independent evaluation of the magnification uncertainties.
The forward-modeling analysis yields effective magnification factors of $\mu_{1a}=5.65^{+0.14}_{-0.65}$ and $\mu_{1bc}=46.8^{+2.3}_{-2.1}$, with uncertainties broadly consistent with the level of variation inferred from the Lenstool reconstruction. We therefore adopt a conservative 10\% uncertainty on the magnification factors throughout this work to account for both lens-model uncertainties and potential differential magnification effects, as listed in Table~\ref{tab:magnification}.

\begin{deluxetable*}
{lccccc}
\setlength{\tabcolsep}{8pt}
\tabletypesize{\scriptsize}
\tablecaption{Model-derived Magnifications \label{tab:magnification}}
\tablehead{
\colhead{Arc} &
\colhead{F200W} &
\colhead{SMA 273 GHz} & \colhead{SMA 225 GHz} &
\colhead{ALMA Band 4} & \colhead{ALMA Band 3}
}
\startdata
Arc~1a & $5.03\pm0.50$ & $5.76\pm0.58$ & $5.74\pm0.57$ & $5.72\pm0.57$ & $5.36\pm0.54$ \\
Arc~1bc & $41.52\pm4.15$ & $50.19\pm5.02$ & $40.31\pm4.03$ & $45.01\pm4.50$ & $43.11\pm4.31$ \\
\enddata
\tablecomments{
Magnification factors were estimated using the best-fitting gravitational lens model of \citet{Patrick2024}, by taking the ratio of the image-plane to source-plane fluxes.
A conservative 10\% uncertainty is adopted for all magnification factors to account for lens-model uncertainties and possible differential magnification effects.
}
\end{deluxetable*}

\section{PHOTOMETRY}  \label{sec:photometry}
In this section, we present the SED fitting analysis of 1a, 1bc and the reconstructed source-plane morphology. We perform both global SED fitting and spatially resolved SED fitting based on Voronoi binning of the PSF-matched JWST/NIRCam data. These analyses aim to constrain the stellar populations, star formation rates (SFRs), and dust attenuation across different spatial scales.\par

\subsection{Global SED Fitting} 
The global SED analysis combines high-resolution JWST/NIRCam imaging with submillimeter continuum data from SMA and ALMA, providing wavelength coverage from the rest-frame ultraviolet to the millimeter regime.
The inclusion of the SMA photometric points improves the constraints on the far-infrared portion of the SED, as these measurements sample much closer to the dust emission peak.
All photometric measurements were performed in a consistent manner by defining apertures in the image plane and mapping them to the reconstructed source plane using the best-fit lens model, thereby correcting for gravitational lensing amplification.\par
For the JWST/NIRCam data, a single elliptical aperture enclosing the full extent of the lensed emission was defined in the image plane and applied consistently across all eight broad-band filters. \par
For the submillimeter continuum data, defining a robust photometric aperture is non-trivial due to the moderate signal-to-noise ratio and extended morphology of the lensed arc. A $3\sigma$ isophotal threshold systematically truncates low-surface-brightness emission, leading to an underestimation of the total flux. Conversely, directly integrating over the $1\sigma$ mask tends to overestimate the flux, as it includes noise-dominated pixels and excludes negative fluctuations.\par
In all cases, the fluxes adopted for the SED analysis were obtained from aperture photometry performed in the reconstructed source plane.
The resulting photometric measurements for the combined system (DSFG-1) are summarized in Table~\ref{tab:photometry}.\par
To obtain a photometric measurement representative of the entire galaxy, we reconstruct and combine the two lensed components (Arc~1a and Arc~1bc) to form the global DSFG-1 system. Arc~1bc traces only a partial region of the galaxy, while Arc~1a contains the full structure but suffers from limited spatial resolution in the SMA and ALMA millimeter data, where the emission corresponding to Arc~1bc is partially blended and diluted. As a result, photometry measured from Arc~1a alone would underestimate the flux in the millimeter bands. By jointly reconstructing Arc~1a and Arc~1bc, we recover a more complete representation of the intrinsic emission of DSFG-1 across all wavelengths.

\begin{deluxetable}{lc}
\setlength{\tabcolsep}{22pt}
\tabletypesize{\scriptsize}
\tablecaption{Source-plane photometry for DSFG-1
\label{tab:photometry}}
\tablehead{
\colhead{Filter} &
\colhead{Flux Density}
}
\startdata
JWST F090W (0.90\,$\mu$m) & $0.359 \pm 0.036\,\mu$Jy \\
JWST F115W (1.15\,$\mu$m) & $0.698 \pm 0.070\,\mu$Jy \\
JWST F150W (1.50\,$\mu$m) & $1.406 \pm 0.141\,\mu$Jy \\
JWST F200W (2.00\,$\mu$m) & $2.557 \pm 0.256\,\mu$Jy \\
JWST F277W (2.77\,$\mu$m) & $3.256 \pm 0.326\,\mu$Jy \\
JWST F356W (3.56\,$\mu$m) & $4.648 \pm 0.465\,\mu$Jy \\
JWST F444W (4.44\,$\mu$m) & $5.912 \pm 0.591\,\mu$Jy \\
JWST F770W (7.70\,$\mu$m) & $6.246 \pm 0.625\,\mu$Jy \\
SMA 1.10\,mm & $441.68 \pm 66.95\,\mu$Jy \\
SMA 1.33\,mm & $226.78 \pm 35.75\,\mu$Jy \\
ALMA 2\,mm & $70.27 \pm 7.52\,\mu$Jy \\
ALMA 3\,mm & $47.52 \pm 5.76\,\mu$Jy \\
\enddata
\tablecomments{Flux densities are measured in the reconstructed source plane.}
\end{deluxetable}

We then fit the SEDs using \texttt{CIGALE} \citep{Boquien2019}, adopting the following modules: \texttt{sfhdelayed} for the star formation history, \texttt{bc03} for the stellar population synthesis, \texttt{nebular} for nebular emission, \texttt{dustatt\_modified\_starburst} for dust attenuation, and \texttt{Dl2014} for dust emission \citep{Dl2007, Dl2014}. 

The best-fit SED models are shown in Figure~\ref{fig:SMA} panels~\textbf{\textit{d)}}--\textbf{\textit{f)}}, corresponding to arcs 1a and 1bc, and the combined system DSFG-1, respectively. The resulting reduced $\chi^2$ values are 0.5 for 1a, 1.1 for 1bc, and 1.2 for DSFG-1, indicating generally good agreement between the models and the observed photometry.
Based on these fits, we then derive the key physical properties including stellar mass, star formation rate (SFR) and dust attenuation.\par
Based on the best-fit SED of the combined system, we derive a stellar mass of $M_{*} = (1.2 \pm 0.4)\times10^{10},M_{\odot}$, a star-formation rate of $\mathrm{SFR} = (103\pm14)\,M_{\odot}\,\mathrm{yr}^{-1}$, a dust attenuation of $A_V = (1.8\pm0.1)$ mag, and an infrared luminosity of $L_{\rm IR}=(9.4\pm0.7)\times10^{11}\,L_{\odot}$.
We note that the infrared luminosity is less securely constrained than the stellar-population properties because the available SMA and ALMA measurements primarily sample the Rayleigh--Jeans side of the dust emission and do not directly probe the peak of the far-infrared SED. Consequently, the inferred $L_{\rm IR}$ remains sensitive to assumptions regarding the dust temperature and lensing magnification, and should therefore be regarded as an approximate estimate of the total dust luminosity.\par
We compare our SED-based SFR with the dust-corrected H$\alpha$ measurements reported in \citet{Frye2024}. In the previous work, Arc~1a and the nearby galaxy NS\_46 were treated as separate components, whereas in our analysis they are modeled as a single system. Although the two SFR indicators are tracing intrinsically different star forming components (integrated over longer time $vs$ ionized gas with shorter timescale),  we find that our SFR ($\sim100\,M_{\odot}\,\mathrm{yr}^{-1}$) is broadly consistent with the combined H$\alpha$-based SFR values for Arc~1a and NS\_46 ($\sim50$--$200\,M_{\odot}\,\mathrm{yr}^{-1}$).
We note that this comparison may also be affected by the different lensing magnification factors adopted,  and are subject to large uncertainties ($\Delta$SFR\,$\sim$15-70\,$M_{\odot}\,\mathrm{yr}^{-1}$).\par
The dust emission is best described by $q_{\mathrm{PAH}} = 3.90$, $U_{\mathrm{min}} = 10.0$, $\alpha = 2.3$, and $\gamma = 0.15$.
These values suggest that the dust is exposed to a relatively intense interstellar radiation field and that a non-negligible fraction of the dust is associated with photodissociation regions, consistent with an actively star-forming dusty galaxy.
The physical properties of the individual arcs (1a and 1bc) are shown separately in Table~\ref{tab:quantity}.\par
Owing to the degeneracy between star-formation history (SFH) and dust attenuation, together with the limited constraints around the Balmer/4000\,\AA\ break region, the mass-weighted stellar age is poorly constrained and is therefore not considered in the following analysis.

%From the best-fit SED for 1a, we derive a stellar mass of $M_{*} = (1.1 \pm 0.4)\times10^{10}\,M_{\odot}$, a star-formation rate of $\mathrm{SFR} = (84.1\pm18.5) \,M_{\odot}\,\mathrm{yr}^{-1}$ and a dust attenuation of $A_{V} = (1.7\pm0.1)$~mag.\par
%From the best-fit SED for 1bc, we derive a stellar mass of $M_{*} = (3.4 \pm 0.5)\times10^{9}\,M_{\odot}$, a star-formation rate of $\mathrm{SFR} = (93.5\pm10.5) \,M_{\odot}\,\mathrm{yr}^{-1}$ and a dust attenuation of $A_{V} = (3.2\pm0.1)$~mag.\par
%From the best-fit SED for DSFG-1, we derive a stellar mass of $M_{*} = (1.2 \pm 0.4)\times10^{9}\,M_{\odot}$, a star-formation rate of $\mathrm{SFR} = (103.4\pm14.1) \,M_{\odot}\,\mathrm{yr}^{-1}$ and a dust attenuation of $A_{V} = (1.8\pm0.1)$~mag.
%The dust emission is best described by $q_{\mathrm{PAH}} = 3.90$, $U_{\mathrm{min}} = 10.0$, $\alpha = 2.3$ and $\gamma = 0.15$.

\subsection{Voronoi-Binned SED Fitting}
\begin{figure*}[t]
    \centering
    \includegraphics[width=\textwidth]{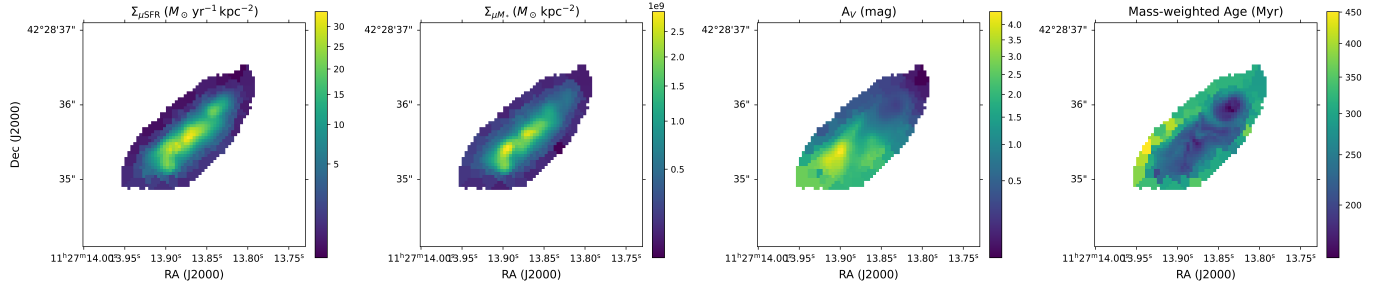}
    \caption{Spatially resolved properties of source 1a in the image plane. The panels show the star formation rate surface density ($\Sigma_{\rm SFR}$), stellar mass surface density ($\Sigma_{M_*}$), dust attenuation ($A_V$), and mass-weighted stellar age. These values are not corrected for gravitational lensing magnification.}
    \label{fig:image_plane_1a}
\end{figure*}

\begin{figure*}[t]
    \centering
    \includegraphics[width=\textwidth]{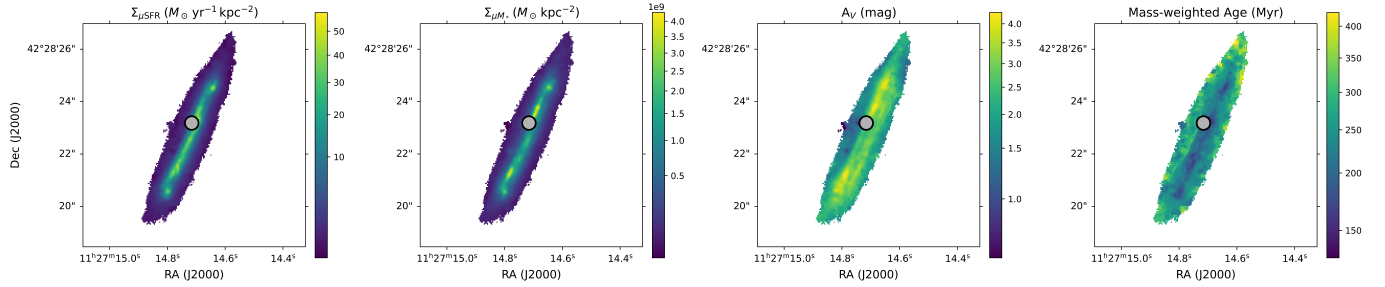}
    \caption{Spatially resolved image-plane properties of Arc~1bc. The panels are the same as in Figure~\ref{fig:image_plane_1a}, but for Arc~1bc. The grey circle indicates the position of a foreground source, which was excluded from the analysis.}
    \label{fig:image_plane_1bc}
\end{figure*}

To investigate the spatial variations in stellar populations and dust attenuation across the lensed system, we performed spatially resolved SED fitting on the Voronoi-binned photometric maps using the eight PSF-matched JWST/NIRCam bands.
Spatial binning was carried out using the adaptive Voronoi tessellation method of \citet{Cappellari2003}, which groups neighboring pixels into irregularly shaped bins to reach a target signal-to-noise ratio (S/N) while preserving the highest possible spatial resolution.
Prior to the spatially resolved analysis, all NIRCam images were convolved to match the F444W point-spread function to ensure consistent aperture colors across filters in the image plane.
%We then constructed S/N maps in each band, adopting the F444W image as the reference for spatial binning.
%A Voronoi tessellation algorithm was applied to the F444W S/N map, with a target S/N threshold of 20 per bin. Each Voronoi bin was fitted individually with CIGALE using the same model setup as the global fitting.
A signal-to-noise (S/N) map was then constructed from the F444W image and used as the reference for the adaptive Voronoi tessellation, with a target S/N of 20 per bin.
Each Voronoi bin was fitted individually with CIGALE using the same model setup as the global fitting.
Given that the NIRCam wavelength coverage is insufficient to independently constrain the dust emission parameters at the spatial-bin level, the dust model parameters (following \citep{Dl2014}) were fixed to the values derived from the global SED fitting to avoid degeneracies caused by the reduced number of photometric points per Voronoi bin and to ensure consistency across the spatially resolved fitting, while the V-band extinction $A_V$ was allowed to vary in each bin.
The spatially resolved physical properties of Arc~1a are presented in Figure~\ref{fig:image_plane_1a}, while those of Arc~1bc are shown in Figure~\ref{fig:image_plane_1bc}.\par 
The reported values represent the surface densities (per kpc$^2$) associated with each pixel rather than the integrated value over the entire Voronoi bin (i.e., the total quantity within each bin has been normalized by the bin area in kpc$^2$). This ensures that the spatial variation of the physical properties is not artificially smoothed by differences in bin area and allows a more direct comparison with the underlying imaging resolution.\par
During the Voronoi binning process, we identified a compact foreground source located immediately in front of Arc~1bc (also see in Figure~\ref{fig:SMA} panel \textbf{\textit{b)}}). The object is clearly resolved in the short-wavelength NIRCam images (F090W–F200W), but becomes indistinguishable in the longer-wavelength bands (F277W–F444W). Its blue color and isolated morphology indicate that it is a foreground galaxy that is not related to the lensed DSFG-1 system. To avoid contamination to the SED fitting of 1bc, this region was masked out and excluded from the Voronoi-binned SED analysis, similarly treated in \citep{Patrick2024} (refer to the grey circle in Figure~\ref{fig:image_plane_1bc}).

\section{Discussion} \label{sec:discussion}
\subsection{Asymmetric dust emission distribution}
\begin{figure}[t]
    \centering
    \includegraphics[width=\columnwidth]{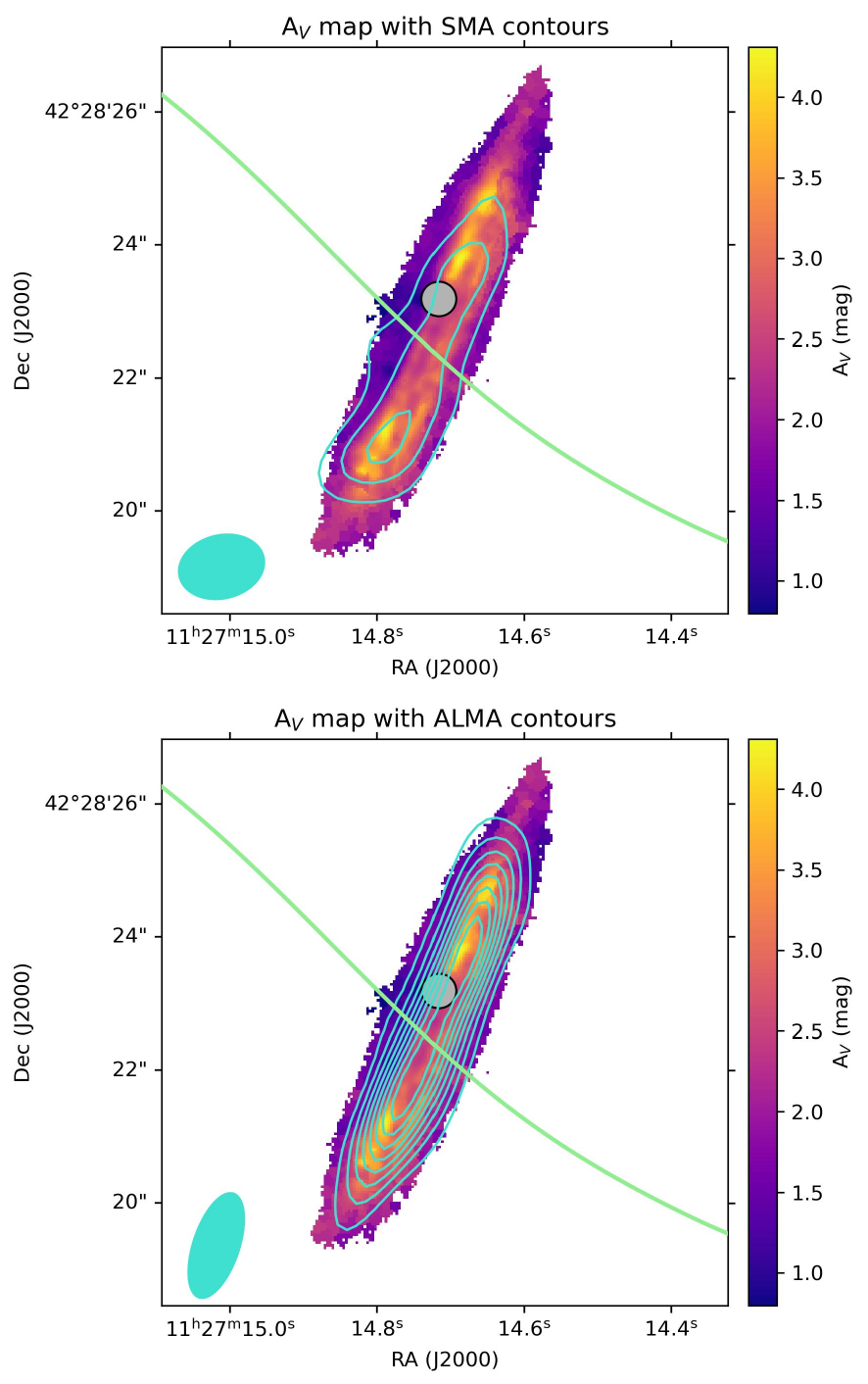}
    \caption{
    %An asymmetric distribution of dust emission revealed by the SMA Rx240+345 observations, shown as cyan contours overlaid on the attenuation ($A_V$) map of component 1bc derived from Voronoi-binned SED fitting. The SMA contours start at $4\sigma$ and increase in steps of $3\sigma$. The grey circle indicates the position of a foreground source. The green curve indicates the lensing critical line in the image plane.
    Comparison of the dust continuum emission and attenuation distribution in Arc~1bc. Upper panel: SMA Rx240+345 dust continuum emission (cyan contours) overlaid on the attenuation ($A_V$) map derived from the Voronoi-binned SED fitting. Lower panel: ALMA dust continuum emission (cyan contours) overlaid on the same $A_V$ map. Contour start at $4\sigma$ and increase in steps of $3\sigma$. The grey circle indicates the position of a foreground source, and the green curve marks the lensing critical line in the image plane.
}
    \label{fig:image_Av_1bc}
\end{figure}

When comparing the SMA continuum morphology with the spatially resolved stellar population maps, we find that the dust continuum emission generally traces regions of enhanced attenuation. In image 1a, the SMA continuum peak is spatially coincident with the highest-$A_V$ region inferred from the Voronoi-binned SED fitting, consistent with the expectation that the millimeter continuum emission originates from the most heavily obscured star-forming component of the galaxy. We note, however, that the resolved $A_V$ values are derived solely from JWST photometry and should therefore be interpreted primarily as indicators of relative attenuation rather than precise measurements of the total dust column density.
A different picture emerges in image 1bc. The optical morphology of images 1b--1c is highly symmetric (upper panel of Figure~\ref{fig:image_Av_1bc}), and this is further confirmed by the spatially binned measurements: the derived physical parameters --- including $A_V$, stellar mass surface density, and SFR --- all display a remarkably symmetric distribution across the lens configuration. In contrast, the SMA continuum contours reveal a clear flux asymmetry between the two images. The south-eastern image~1c shows a significantly stronger sub-mm continuum detection (up to $3\sigma$) than the north-western counter-image~1b. Such asymmetry was not observed in previous ALMA measurements (lower panel of Figure~\ref{fig:image_Av_1bc}), where the two images appeared consistent in flux and morphology within uncertainties.\par
\begin{figure*}[t]
    \centering
    \includegraphics[width=\textwidth]{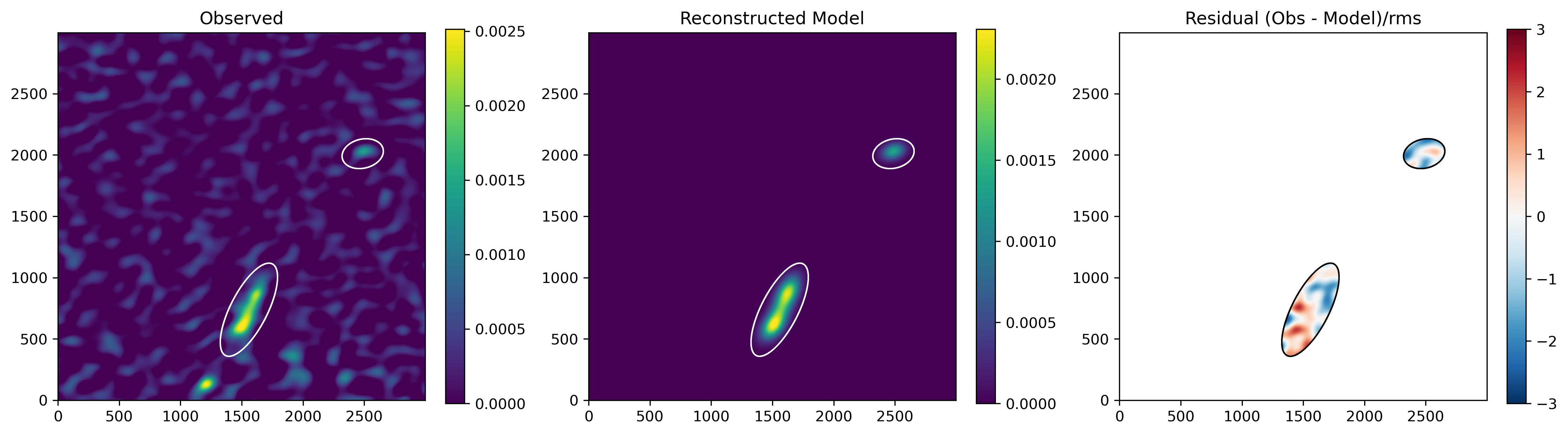}
    \caption{
    Forward-modeling best-fit result of the SMA 240+345 GHz continuum emission. The left panel shows the observed SMA image-plane continuum map, the middle panel shows the reconstructed image-plane model generated from a symmetric source-plane Sérsic profile and propagated through the lens model, and the right panel shows the nomalized residual map, $(\mathrm{obs}-\mathrm{model})/\mathrm{rms}$, which quantifies the significance of the residuals relative to the local noise level. The model successfully reproduces the apparent flux asymmetry between images 1b and 1c without requiring intrinsic asymmetry in the source. The normalized residuals remain below $3\sigma$ across the fitted region, indicating that the observed morphology is well explained by the combined effects of differential magnification and the anisotropic SMA synthesized beam.
}
    \label{fig:test}
\end{figure*}

%When comparing the SMA continuum morphology with the spatially resolved stellar population maps (see Figure~\ref{fig:image_Av_1bc}), the system appears notably inconsistent with the expected symmetry. The optical structural morphology of images 1b–1c is highly symmetric, and this is further confirmed by the spatially binned measurements: the derived physical parameters — including $A_V$, stellar mass surface density, and SFR — all display a remarkably symmetric distribution across the lens configuration. In contrast, the SMA continuum contours reveal a clear flux asymmetry between the two images. The south-estern image (1c) shows a significantly stronger sub-mm continuum detection (up to $3\sigma$) than the north-weastern counter-image (1b). Such asymmetry was not observed in previous ALMA measurements, where the two images appeared consistent in flux and morphology within uncertainties.\par
%A plausible explanation lies in the instrumental response. The ALMA synthesized beam is closely aligned with the major symmetry axis of the 1bc configuration, potentially smoothing or suppressing intrinsic sub-structure due to beam elongation. In contrast, the SMA beam is oriented at a substantially different position angle, providing sensitivity to spatial variations along the direction where 1b and 1c are separated. This geometric difference may allow SMA to resolve brightness variations that were effectively beam-averaged in the ALMA data.\par
To investigate this discrepancy, we perform forward modeling of the dust continuum emission (see Section~\ref{sec:Re} for more information).
We find that the observed flux asymmetry between images 1b and 1c can be naturally reproduced without invoking intrinsic differences in the source (see Figure~\ref{fig:test}). Instead, the effect arises from differential magnification coupled with the elliptic SMA beam, which selectively enhances brightness variations along the direction separating the two images. In particular, small-scale structure in the source plane is stretched differently across the critical curve, leading to unequal flux recovery once convolved with the interferometric beam. This effect is not apparent in the ALMA data, where the synthesized beam is more closely aligned with the symmetry axis of the lens configuration, effectively averaging over these variations.
More importantly, the ability of the forward model to simultaneously reproduce both the morphology and the relative fluxes of images 1b and 1c provides independent validation of the adopted lens model. We therefore conclude that the apparent asymmetry in the SMA continuum does not reflect intrinsic differences in the source, but instead arises from the combined effects of lensing geometry and instrumental response.
We note that the foreground source located in the northern region of Arc~1bc (see Figure~\ref{fig:image_plane_1bc}) lies in close proximity to image 1b and may introduce a minor contribution to the observed SMA continuum emission, potentially affecting the flux distribution at a low level. However, this effect is expected to be small and does not impact our main conclusions.

\subsection{Resolved vs. global properties}
When comparing the global and resolved measurements, which is presented in Table \ref{tab:quantity}, we find that the globally inferred SFR is significantly lower than the spatially resolved SFR by a factor of $\sim1.5$. This discrepancy is primarily driven by the lack of long-wavelength constraints in the resolved SED fitting: while the global SED includes the ALMA and SMA continuum measurements, the resolved SED fitting relies solely on JWST/NIRCam data and therefore lacks the far-infrared/millimeter coverage that directly traces dust-attenuated star formation. As a result, the resolved SFRs tend to overestimate the current star-formation activity due to stronger age–dust–SFR degeneracies and the inability to constrain the obscured component of the star formation without long-wavelength data.\par
Previous spatially resolved studies have suggested that integrated SED fitting may underestimate stellar masses relative to resolved approaches \citet{Sorba2015,Sorba2018}. While our current results do not show evidence of such a discrepancy, we emphasize that this conclusion is based on a small sample and may not be generalizable.

\begin{deluxetable*}
{lcccc}
\tabletypesize{\scriptsize}
\tablewidth{0pt}
\tablecaption{Comparison between Global and Spatially Resolved SED Fitting Results \label{tab:quantity}}
\tablehead{
\colhead{Arc} &
\colhead{Quantity} &
\colhead{Global SED} &
\colhead{Resolved SED (Voronoi sum)} &
\colhead{Resolved / Global}
}
\startdata
Arc~1a & SFR [$M_\odot\,\mathrm{yr}^{-1}$] & $84 \pm 19$ & $123 \pm 5$ & $1.5 \pm 0.3$ \\
Arc~1a & $M_\star$ [$10^{10}\,M_\odot$]    & $1.1 \pm 0.5$ & $1.2 \pm 0.1$ & $1.1 \pm 0.5$ \\
Arc~1a & Av [mag] & $1.7\pm0.1$ & $2.3\pm0.3$ & $1.4\pm0.2$\\
\hline
Arc~1bc & SFR [$M_\odot\,\mathrm{yr}^{-1}$] & $94 \pm 11$ & $69 \pm 3$ & $0.7 \pm 0.1$ \\
Arc~1bc & $M_\star$ [$10^{9}\,M_\odot$]    & $3.4 \pm 0.5$ & $5.6 \pm 0.1$ & $1.6 \pm 0.2$ \\
Arc~1bc & Av [mag] & $3.2\pm0.1$ & $2.6\pm0.4$ & $0.8\pm0.1$\\
\enddata
\tablecomments{
The global SED fitting includes JWST/NIRCam imaging together with ALMA and SMA continuum measurements, while the spatially resolved SED fitting relies solely on JWST/NIRCam data. The resolved values are obtained by summing the Voronoi-binned, per-pixel physical quantities over the entire source. All physical quantities used in this comparison have been corrected for the magnification factor.}
\end{deluxetable*}

\subsection{Is DSFG-1 a merger?}
\begin{figure*}[t]
    \centering
    \includegraphics[width=\textwidth]{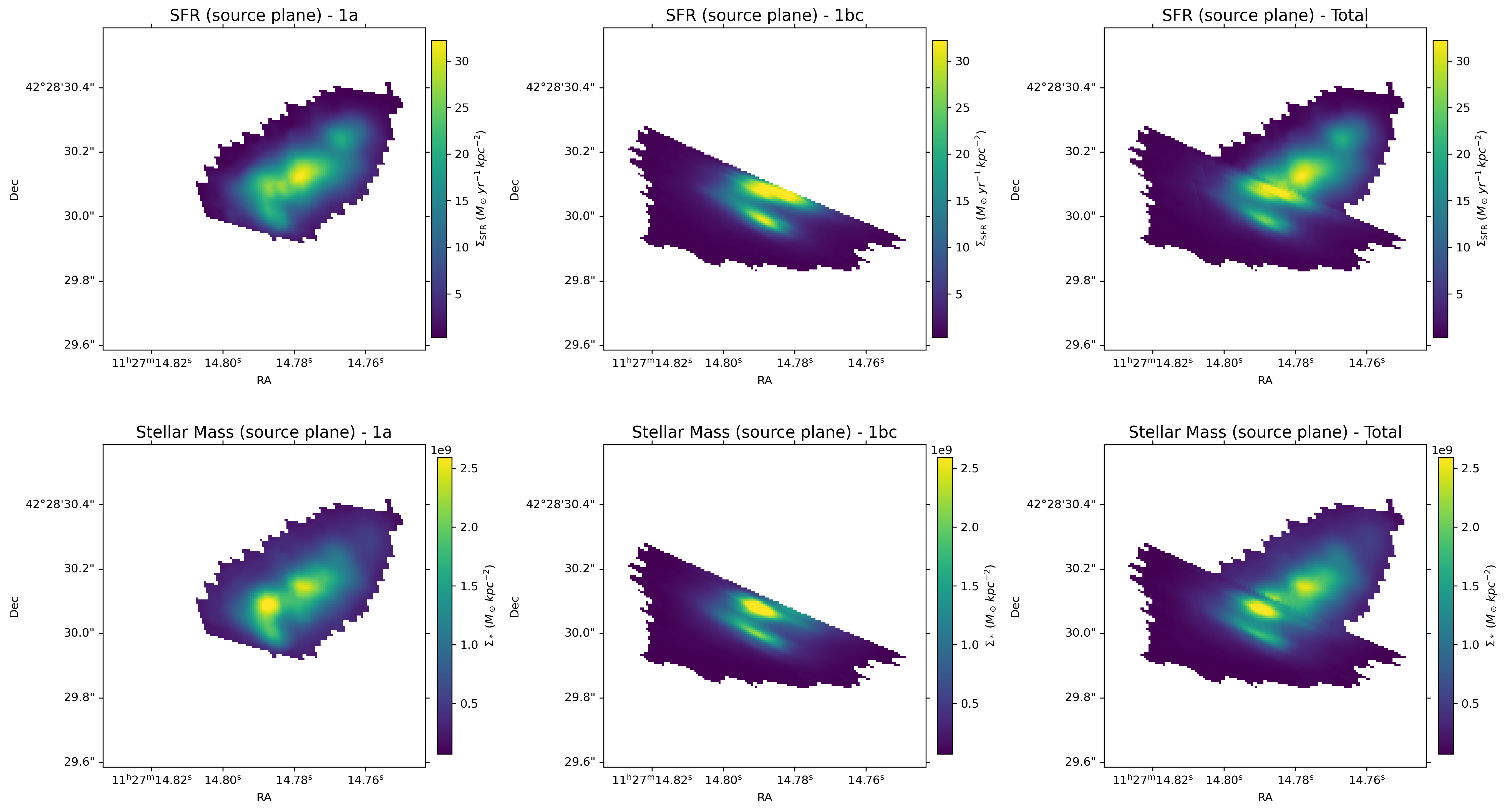}
    \caption{Resolved source-plane surface density maps of G165 DSFG-1 derived from the spatially resolved SED fitting. The top row shows the star formation rate surface density ($\Sigma_{\rm SFR}$), while the bottom row presents the stellar mass surface density ($\Sigma_{\rm M_*}$). From left to right, the panels show the independent source-plane reconstructions derived from image~1a, image~1bc, and the combined reconstruction using both lensed images.
    The stellar mass reconstructions consistently recover a bimodal morphology with two dominant mass concentrations separated by $\sim1.3$~kpc. In contrast, the SFR reconstructions exhibit more complex and reconstruction-dependent substructures, likely reflecting the combined effects of differential magnification, lensing geometry, and limited spatial resolution. The highly magnified 1bc reconstruction provides stronger constraints on the extended morphology of the system. See Section~4.3 for further discussion.}
    \label{fig:source_prop}
\end{figure*}

To assess whether G165 DSFG-1 is undergoing a merger event, we reconstruct the spatially resolved physical properties into the source plane with both lensed images 1a and 1bc. 
Starting from the image-plane Voronoi-binned SED fitting results, we map the derived physical quantities back to the source plane using the best-fitting gravitational lens model, ensuring that lensing distortions and spatial magnification variations are properly taken into account.
Figure~\ref{fig:source_prop} presents the independent source-plane reconstructions derived from image~1a, image~1bc, and their combined reconstruction, allowing a direct comparison of the inferred morphologies from the different lensed images.\par
%The source-plane stellar mass map (left panel of Figure~\ref{fig:source_prop}) reveals two distinct mass concentrations separated by $0\farcs157$, corresponding to $\sim1.33$kpc at $z=2.236$. This result is robust against the details of the lensing reconstruction. The presence of two stellar mass peaks is already apparent in the source-plane reconstruction based solely on image 1a, and remains qualitatively unchanged when both image 1a and 1bc are jointly reconstructed. While the relative prominence of the two components and the detailed morphology vary slightly between the two reconstructions, the overall bimodal mass distribution and their projected separation are preserved.\par
%The source-plane SFR distributions map (right panel of Figure~\ref{fig:source_prop}) shows a more complex behavior. In the reconstruction based on image 1a alone (not shown), the SFR map exhibits multiple peaks located between the two stellar mass components, with one peak lying closer to the right-hand stellar mass concentration. In contrast, the joint 1a and 1bc reconstruction, as shown in Fig~\ref{fig:source_prop}, yields a more centrally concentrated SFR distribution, preferentially aligned with the left stellar mass peak, and with a noticeable difference in peak intensity. These discrepancies likely reflect the combined effects of lensing geometry, differential magnification, and the limited spatial resolution of the SED-based SFR maps, and therefore should be interpreted with caution.\par
The source-plane stellar mass maps (bottom row of Figure~\ref{fig:source_prop}) reveal two distinct mass concentrations separated by $\sim0\farcs157$, corresponding to $\sim1.33$~kpc at $z=2.236$. Importantly, this bimodal morphology is independently recovered in both the 1a-only and 1bc-only reconstructions, although the relative prominence and detailed morphology of the two components vary slightly between the two images. The combined reconstruction preserves the overall two-component structure, suggesting that the inferred stellar mass distribution is not driven solely by one particular lensed image.
The reconstructed SFR surface density maps (top row of Figure~\ref{fig:source_prop}) exhibit a more complex morphology than the stellar mass distributions. The 1a-only reconstruction shows a dominant central peak together with an additional weaker peak, whereas the 1bc-only reconstruction is characterized primarily by a single strong elongated component. In the combined reconstruction, both structures contribute to the final morphology, resulting in two main SFR peaks distributed across the central region of the system.
These differences likely reflect the combined effects of differential magnification, lensing geometry, and the limited spatial resolution of the reconstructed SFR maps. In particular, the highly stretched 1bc image provides tighter constraints on the extended structure near the critical curve, while the 1a reconstruction contributes additional information on the secondary emission component. In addition, because the spatially resolved SED fitting is constrained solely by the JWST/NIRCam photometry, unresolved heavily obscured stellar populations or star-forming regions may introduce additional uncertainty into the inferred stellar masses and SFRs in the dustiest parts of the system. Therefore, while the detailed small-scale SFR morphology remains somewhat reconstruction-dependent, the overall picture suggests a spatially extended and multi-component star-forming structure.
Together, based on the current results, G165 DSFG-1 shows compelling evidence for a multi-component system, with source-plane stellar mass reconstructions revealing two distinct mass concentrations. These features are suggestive of a merger scenario; however, the present data are insufficient to unambiguously classify the system as a major merger.\par
Our interpretation is also supported by previous spectroscopic evidence from \citet{Frye2024}, which identified Arc~1 as an interacting galaxy pair based on JWST/NIRSpec observations. In their study, the two components (Arc~1a and NS\_46) exhibit a velocity separation of $\sim420\,\mathrm{km\,s^{-1}}$ and a projected separation of $\sim5\,\mathrm{kpc}$, providing direct evidence for an ongoing interaction. While our analysis is based on spatially resolved SED modeling and lensing reconstruction, the spectroscopic measurements offer complementary kinematic constraints. These independent observational results, combined with our resolved stellar component distribution,  consistently support a merger interpretation of the system.\par
Notably, there is a nearby galaxy, DSFG-3, which is also detected in our SMA data, located at the same redshift but only visible in image Arc~3c, since its dusty component does not extend into the region covered by Arc~3ab \citep[see also][]{Patrick2024}. The separation between DSFG-1 and DSFG-3 in the source plane is less than 10~kpc, suggesting that they have likely interacted or may do so in the near future. In addition, the multiple stellar mass peaks within DSFG-1 itself may indicate yet another recent interaction. Further observations will be essential to confirm the merger nature and to characterize its dynamical state.

\subsection{Comparison with SMGs in the literature}
\label{sec:Re}
%Due to the complexity in the lensing reconstruction and beam convolution in the source-plane images, the effective radius (Re) is estimated based on the number of pixels encountered within the peak 50\% of the reconstructed flux and should therefore be regarded as an upper limit ($R_\mathrm{e}\lesssim2.4$~kpc).\par

To obtain a more robust constraint on the intrinsic dust-emitting size, we adopt a forward-modeling approach rather than relying directly on measurements in the reconstructed source plane. We model the SMA continuum emission using a Sérsic profile parameterized by an effective radius $R_\mathrm{e}$, Sérsic index $n$, axis ratio $q$, and position angle $PA$. The intrinsic model is projected into the image plane through the lens model, convolved with the SMA synthesized beam, and directly compared with the observed data using an MCMC framework. This approach naturally accounts for beam smearing and lensing distortion, which can otherwise bias source-plane size estimates derived from reconstructed images. The posterior distribution favors a characteristic source-plane effective radius of $R_\mathrm{e}=1.1^{+0.5}_{-0.4}$~kpc. The corresponding posterior distributions and parameters are presented in Appendix~\ref{sect:mcmc}.\par
\begin{figure*}[ht!]
    \centering
    \includegraphics[width=0.9\linewidth, trim=0 0 0 0]{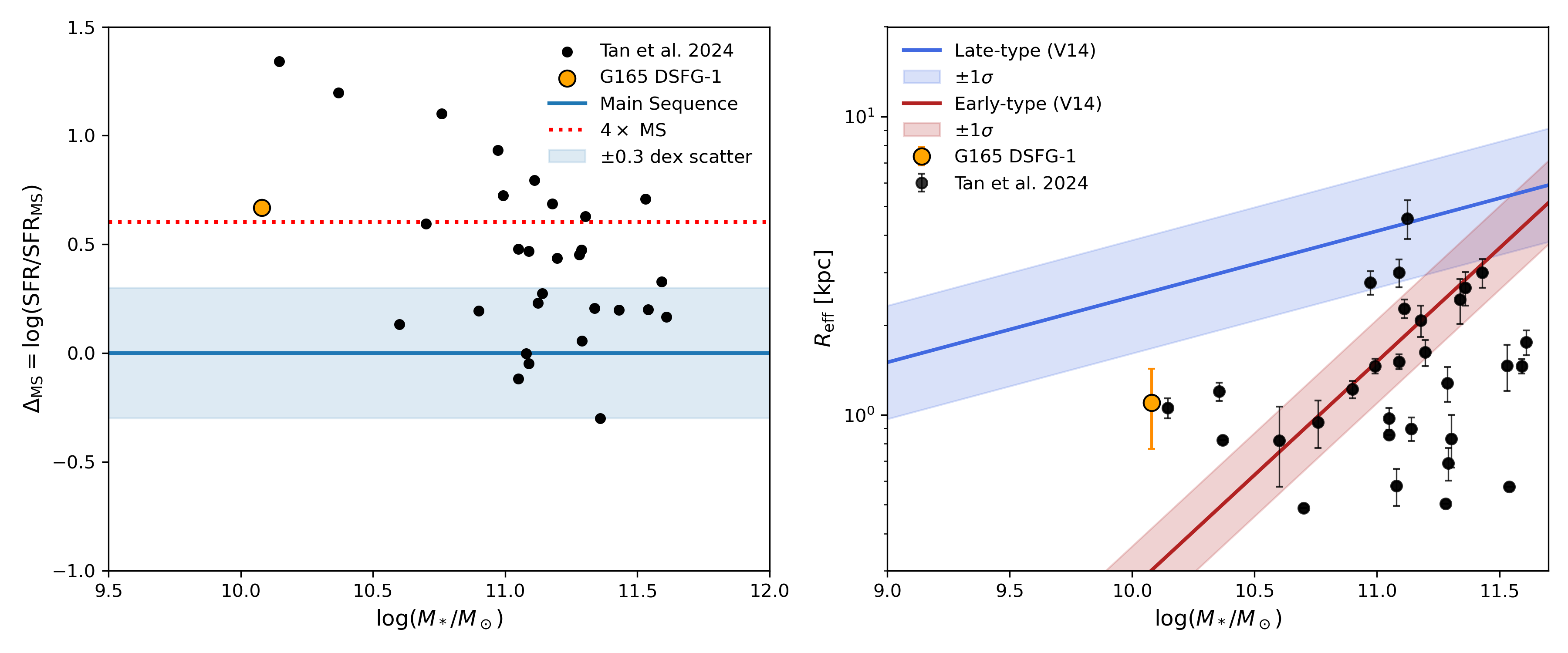}

    \caption{
    Left panel: Offset from the star-forming main sequence ($\Delta_{\rm MS} = \log (\mathrm{SFR}/\mathrm{SFR}_{\mathrm{MS}})$) as a function of stellar mass for G165 DSFG-1. Black circles represent the comparison sample selected from \citet{Tan2024} in the redshift range $1.8 < z < 2.8$. The orange circle indicates G165 DSFG-1. The horizontal blue line marks the main sequence ($\Delta{\rm MS}=0$) based on the parameterization of \citet{Schreiber2015}. The shaded region indicates the intrinsic scatter of $\pm0.3$ dex around the main sequence. The red dashed line denotes the starburst threshold at four times the main-sequence star formation rate.
    Right panel: Stellar mass–size relation for G165 DSFG-1. The black points show comparison galaxies from \citet{Tan2024}, selected in the redshift range $1.8 < z < 2.8$. The orange circle marks G165 DSFG-1. The blue and red solid lines denote the late-type and early-type size–mass relations at $z=2.25$ from \citet{van2014}, with the shaded regions indicating the corresponding $1\sigma$ intrinsic scatter.
    \label{fig:discussion}
    }
\end{figure*}

%This approach naturally accounts for the effects of lensing distortion and beam smearing, which can otherwise bias size estimates derived from reconstructed images. We obtain an effective radius of $R_\mathrm{e} = 1.6 \pm 0.4$~kpc (source plane), along with constraints on the morphology ($q$, PA).
To place this system in a broader evolutionary context, we compare its global star-forming and structural properties with those of distant submillimeter-bright galaxies undergoing in-situ spheroid formation \citep{ Tan2024}.
As shown in Figure~\ref{fig:discussion}, G165~DSFG-1 lies at the lower-mass end of the SMG population, with a stellar mass of $\sim10^{10}M_\odot$ \citep{Gillman2024}. This identifies it as a relatively less massive SMG undergoing an intense starburst phase, while occupying a distinctive yet physically meaningful position in both the star formation rate–stellar mass and size–mass planes compared to the comparison sample.\par
In the SFR–$M_*$ diagram (left panel in Figure~\ref{fig:discussion}), G165~DSFG-1 is located $\sim$4 times above the star-forming main sequence at $z\simeq2.236$, indicating enhanced star formation activity comparable to that observed in submillimeter-selected systems associated with rapid, in-situ spheroid growth.
In the $R_\mathrm{e}$--$M_*$ plane (right panel in Figure~\ref{fig:discussion}), G165 DSFG-1 occupies an intermediate region between the late-type and early-type size-mass relations at similar redshift. The forward-modeling analysis yields an intrinsic dust-continuum effective radius of $R_\mathrm{e}=1.1^{+0.5}_{-0.4}$~kpc, indicating a compact dust-emitting structure comparable to those observed in submillimeter-selected galaxies undergoing rapid stellar mass assembly \citep[e.g.][]{Tan2024}.\par
This compact size, together with the elevated star formation rate, is consistent with a scenario in which star formation is concentrated within a relatively small physical region. However, the uncertainty on the inferred size remains non-negligible owing to the limited angular resolution of the SMA observations and the assumptions adopted in the forward-modeling procedure.\par
A similar contrast is also evident when extending the comparison to dusty star-forming galaxies at lower redshift. Morphological studies of $z\sim1$ ULIRGs selected at 16~$\mu$m \citep{Liang2024} show that, at comparable infrared luminosities, a substantial fraction of ULIRGs display compact sizes and bulge-dominated morphologies, with roughly half of the ULIRG population lying close to the early-type size–mass relation. These lower-redshift ULIRGs are typically more massive, more centrally concentrated, and often associated with enhanced Active Galactic Nucleus (AGN) activity.
In contrast, G165~DSFG-1 retains a less centrally concentrated morphology.
This difference suggests that intense infrared-luminous star formation at high redshift does not necessarily coincide with immediate structural compaction, and that the evolutionary pathways leading to compact early-type systems may depend on redshift, stellar mass, and the timing or nature of dynamical processes such as mergers or dissipative collapse.\par
From an evolutionary perspective, the combination of elevated star formation activity and moderately compact structure may indicate that G165~DSFG-1 is undergoing a phase of rapid stellar mass assembly associated with structural transformation.
At $z=2.236$, the age of the Universe is approximately $\sim$3~Gyr (for $H_0=70~\mathrm{km\,s^{-1}\,Mpc^{-1}}$). If the system were to sustain its current level of star formation for $\sim$0.5~Gyr without significant size growth, it would reach a stellar mass and structural state comparable to early-type galaxies observed at $z\sim1.8$ \citep[e.g.][]{Toft2014}.
In this scenario, G165~DSFG-1 may plausibly represent a transitional phase between actively star-forming systems and more compact spheroid-dominated galaxies. However, we caution that the inferred evolutionary pathway remains uncertain, particularly given the potential effects of differential magnification, beam smearing, and possible future merger or quenching processes.

\section{Summary and Conclusions}
We have presented a detailed study of the strongly lensed dusty star-forming galaxy G165 DSFG-1 at $z=2.236$, combining JWST/NIRCam imaging with Submillimeter Array (SMA) continuum observations.
Strong gravitational lensing produces multiple images with magnification factors ranging from $\mu\sim5$ to $\mu\sim40$, enabling a spatially resolved investigation of the stellar populations and dust-attenuated star formation on sub-kiloparsec scales.
From integrated spectral energy distribution modeling, we find that G165 DSFG-1 has a total star-formation rate (SFR) of $\sim103.4\,M_{\odot}\,\mathrm{yr^{-1}}$ and a stellar mass of $M_{\star} = (1.2 \pm 0.4) \times 10^{10}\,M_{\odot}$, placing it at the lower-mass end of the submillimeter galaxy population and identifying it as a less massive SMG. At this redshift, the system lies about four times above the star-forming main sequence.
Despite this, its compact morphology, with an effective radius of $R_{\rm e}\sim1.1\,\mathrm{kpc}$, and strong dust continuum emission are characteristic of compact starburst systems at $z\sim2.236$.
Using Voronoi-binned, spatially resolved SED fitting, we uncover pronounced internal variations in stellar age and dust attenuation, 
%indicating a non-uniform star-formation history.
indicating that the stellar populations and dust distribution are not uniform across the galaxy.
%These gradients highlight the importance of spatially resolved analyses for interpreting highly magnified systems, particularly in the presence of differential gravitational magnification.
These results highlight the value of spatially resolved studies for probing the internal properties of high-redshift galaxies and demonstrate that significant variations in stellar populations and dust attenuation can exist on sub-kiloparsec scales.
The SMA continuum image reveals strong submillimeter dust emission associated with the highly magnified Arc~1bc.
Combining 225 and 273 GHz continuum data, we identify an asymmetry in the dust continuum morphology of 1bc that differs from the structure inferred at other wavelengths.
The origin of this discrepancy can be interpreted as a combination of differential lensing and beam smearing, which together modify the apparent surface-brightness distribution in the image plane and produce an artificial flux asymmetry between images 1b and 1c, even when the intrinsic source structure is symmetric.
Taken together, our spatially resolved analysis reveals a complex, multi-component structure. These results, combined with previous spectroscopic evidence, are consistent with a merger-driven scenario, although the current data do not yet allow a definitive classification of its dynamical state. 
The compact size, high stellar mass density, and substantial obscured star formation of G165 DSFG-1 suggest that it is undergoing rapid stellar mass assembly and may be in the process of transitioning toward an early-type galaxy. 
%This work illustrates the power of combining gravitational lensing with JWST imaging to probe the internal structure and evolutionary state of dusty galaxies at high redshift, and provides a pathfinder for future spatially resolved morphological and kinemetical studies in the JWST era.
This work highlights the synergy between high-resolution JWST imaging, submillimeter interferometric observations, and strong gravitational lensing in revealing the spatially resolved properties of dusty star-forming galaxies at high redshift, providing new insights into the interplay between stellar populations, dust attenuation, and obscured star formation on sub-kiloparsec scales.

\section{Acknowledgement}
We thank the anonymous referee for the careful reading and constructive suggestions.
This work is sponsored by the National Key R\& D Program of China for grant No. \ 2022YFA1605300, 
the National Natural Science Foundation of China (NSFC) grants No.\ 12273051.
Support for this work is also partly provided by the CASSACA.
P.S.K. acknowledges financial support from the Knut and Alice Wallenberg Foundation.
Q.T. acknowledges support from the NSFC (grant No. 12573013).
DL acknowledges the support from the Strategic Priority Research Program of the Chinese Academy of Sciences, grant No. XDB0800401.\par
This work is based on observations made with the NASA/ESA/CSA James Webb Space Telescope. 
These observations are associated with JWST program \#1176 (PEARLS; PI: R. Windhorst).
The JWST data used can be accessed via \doi{10.17909/551m-fd29}.\par
This paper makes use of the following ALMA data: ADS/JAO.ALMA\#2021.1.00607.S. ALMA is a partnership of ESO (representing its member states), NSF (USA) and NINS (Japan), together with NRC (Canada), NSTC and ASIAA (Taiwan), and KASI (Republic of Korea), in cooperation with the Republic of Chile. The Joint ALMA Observatory is operated by ESO, AUI/NRAO and NAOJ.
The Submillimeter Array is a joint project between the Smithsonian Astrophysical Observatory and the Academia Sinica Institute of Astronomy and Astrophysics and is funded by the Smithsonian Institution and the Academia Sinica.\par
Some of the data products presented herein were retrieved from the Dawn JWST Archive (DJA). DJA is an initiative of the Cosmic Dawn Center (DAWN), which is funded by the Danish National Research Foundation under grant DNRF140. The imaging data products were reduced using the \texttt{grizli} pipeline \citep{grizli2018}, with basic details of the NIRCam data reduction described in \citet{Valentino2023}.\par
\facilities{JWST (NIRCam), SMA, ALMA}
\software{Astropy \citep{Astropy2013, astropy2018, astropy2022}, CASA \citep{McMullin2007}, Matplotlib \citep{Matplotlib2007}, Numpy \citep{Numpy2020}, Scipy \citep{Scipy2020}, VorBin \citep{Cappellari2003}}

\bibliography{sample701}{}
\bibliographystyle{aasjournalv7}  

\clearpage
\appendix
\label{sect:app}
\section{Comparison with ALMA Continuum Imaging} \label{sect:ALMA}
\begin{figure}[t]
    \centering
    \includegraphics[width=\columnwidth]{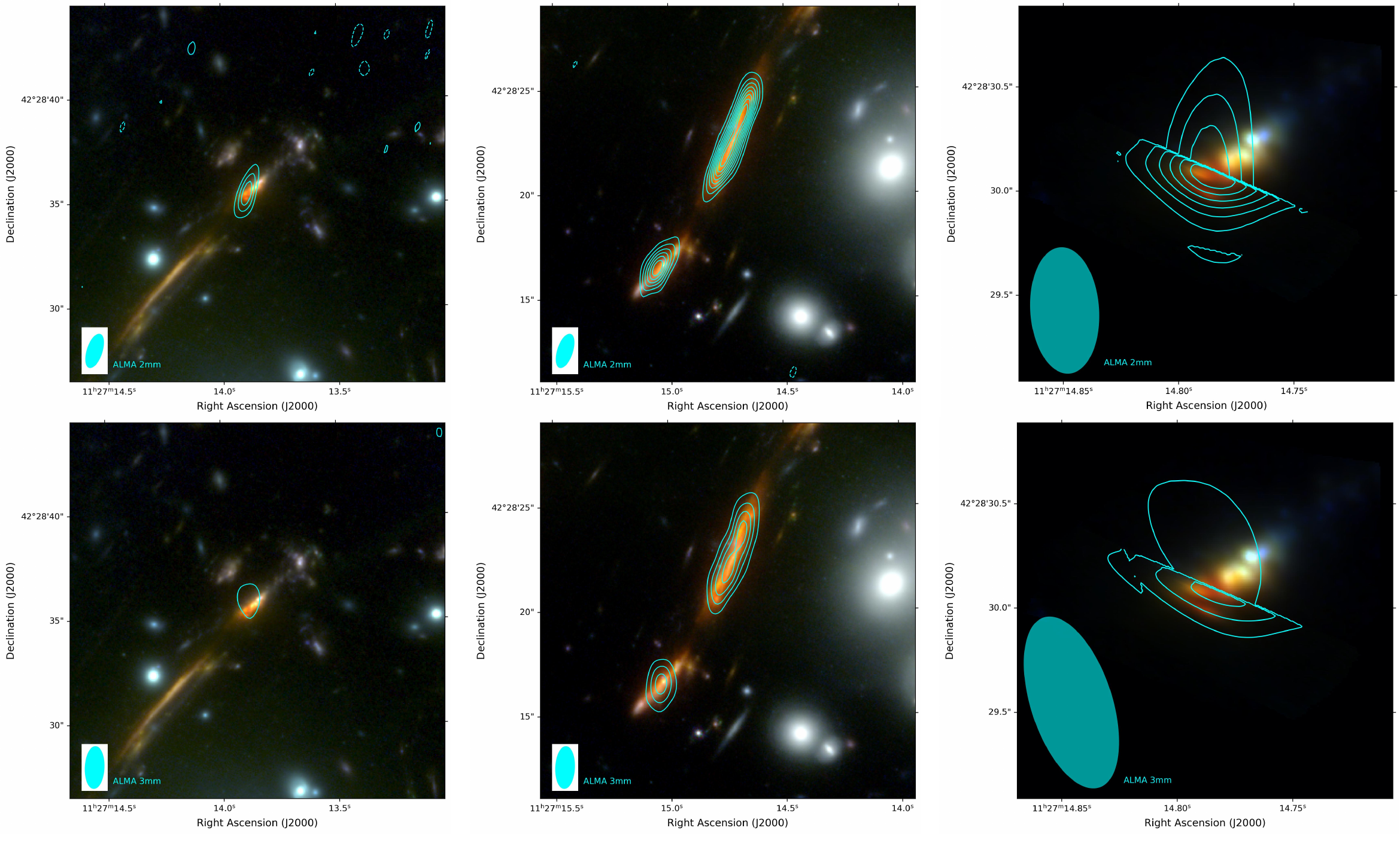}
    \caption{
    JWST NIRCam RGB images (F115W, F277W, and F444W) overlaid with ALMA continuum emission contours.
    Panels \textbf{\textit{a}} and \textbf{\textit{b}} show the image-plane JWST RGB images of Arc~1a and Arc~1bc, respectively, with ALMA 2\,mm continuum contours overlaid. Panel \textbf{\textit{c}} shows the reconstructed source-plane JWST RGB image of DSFG-1 system with the corresponding reconstructed 2\,mm continuum contours.
    Panels \textbf{\textit{d}}--\textbf{\textit{f}} show the same regions and reconstruction as panels \textbf{\textit{a}}--\textbf{\textit{c}}, but for the ALMA 3\,mm continuum emission.
    Contours are shown at $-4\sigma$ (dashed) and positive levels beginning at $4\sigma$ with $3\sigma$ intervals (solid).
    The source-plane beam is derived by projecting the image-plane ALMA beams of arcs 1a and 1bc through the gravitational lensing model and averaging the resulting source-plane beams.
}
    \label{fig:contour_ALMA}
\end{figure}

To provide additional context for the long-wavelength continuum emission, Figure~\ref{fig:contour_ALMA} shows archival ALMA 2 and 3 mm continuum maps overlaid on the JWST images in both the image plane and the reconstructed source plane.\par
The ALMA continuum emission is detected toward the same highly reddened component identified by the SMA observations. The overall spatial distribution is consistent with the SMA continuum morphology, indicating that the dust-obscured star formation is concentrated in the southeastern region of the system.\par

\section{Posterior Distributions of the Forward-model Parameters}
\begin{figure}[t]
    \centering
    \includegraphics[width=\columnwidth]{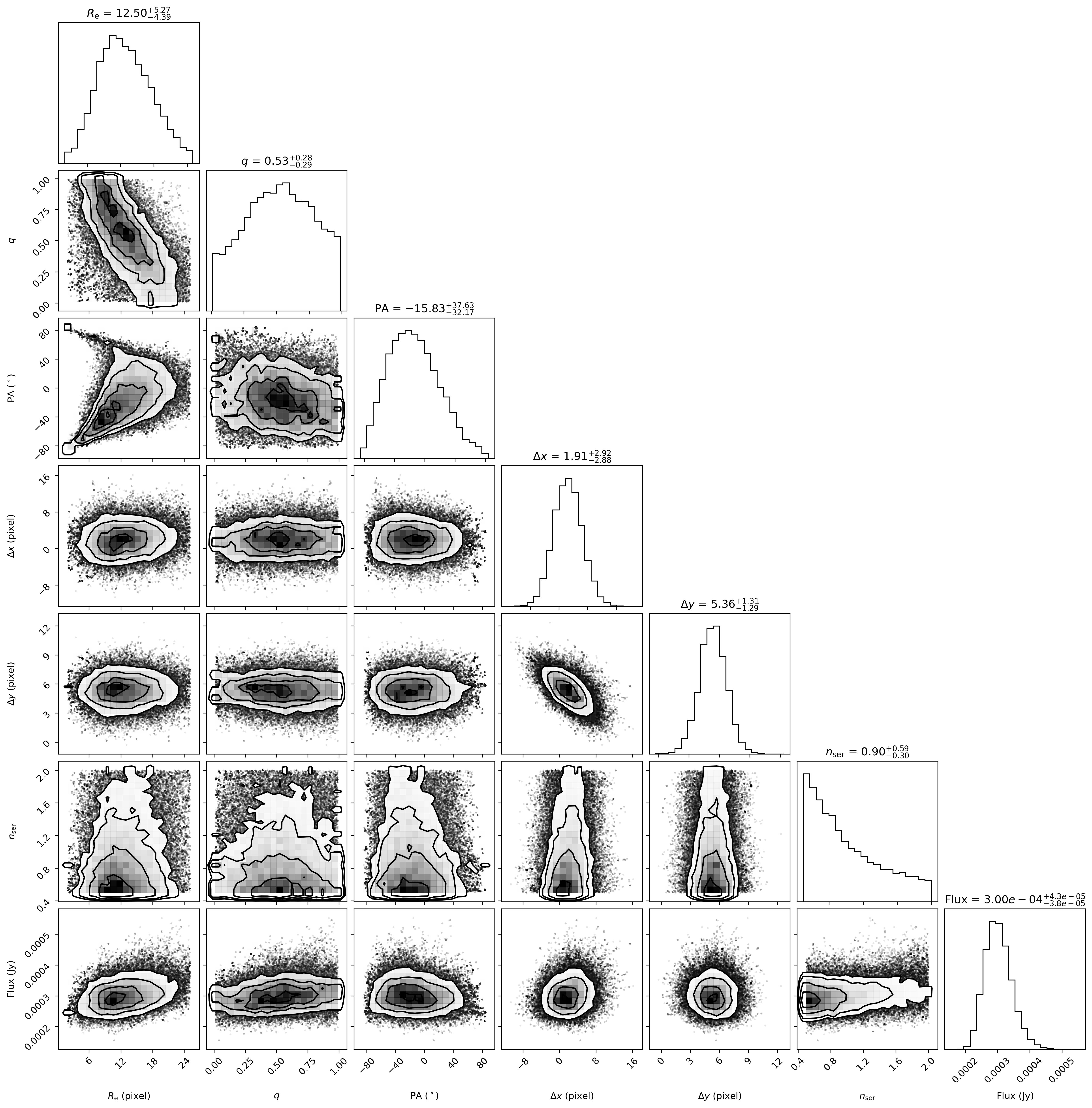}
    \caption{
    Posterior distributions of the forward-model parameters obtained from the MCMC analysis. The dashed lines indicate the 16th, 50th, and 84th percentiles of the marginalized posterior distributions. Parameters include the effective radius ($R_{\rm e}$), axis ratio ($q$), position angle (PA), source-plane position offsets ($dx$, $dy$), Sérsic index ($n$) and intrinsic flux density.
}
    \label{fig:mcmc}
\end{figure}

The posterior distributions of the forward-model parameters described in Section~\ref{sec:Re} are shown in Figure~\ref{fig:mcmc}.\par
The inferred effective radius is $R_{\rm e}=12.5^{+5.3}_{-4.4}$ pixels, corresponding to $R{\rm e}=1.1^{+0.5}_{-0.4}$,kpc assuming a source-plane pixel scale of $0.01''$\,pix$^{-1}$ (0.085\,kpc\,pix$^{-1}$ at $z=2.236$). The posterior distributions indicate that the effective radius is reasonably constrained, while moderate degeneracies remain among several structural parameters owing to the limited angular resolution and signal-to-noise ratio of the SMA observations.
\label{sect:mcmc}

\begin{comment}
\begin{deluxetable*}{lcccc}
\tabletypesize{\scriptsize}
\setlength{\tabcolsep}{12pt}
\tablewidth{0pt}
\tablecaption{Posterior Constraints from the SMA Forward-modeling Analysis
\label{tab:mcmc}}
\tablehead{
\colhead{Parameter} &
\colhead{Unit} &
\colhead{Median} &
\colhead{16th percentile} &
\colhead{84th percentile}
}
\startdata
$R_{\rm e}$ & pix & 12.57 & 8.19 & 17.76 \\
$q$ &  -- & 0.53 & 0.24 & 0.81 \\
PA & deg & -16.65 & -48.21 & 20.95 \\
$dx$ & pix & 2.01 & -0.93 & 4.99 \\
$dy$ & pix & 5.35 & 4.04 & 6.69 \\
$n_{\rm ser}$ & -- & 0.91 & 0.61 & 1.49 \\
Flux & Jy & $3.02\times10^{-4}$ & $2.64\times10^{-4}$ & $3.46\times10^{-4}$ \\
$\mu_{\rm 1a}$ & -- & 5.65 & 5.00 & 5.79 \\
$\mu_{\rm 1bc}$ & -- & 46.75 & 44.66 & 49.00 \\
\enddata
\end{deluxetable*}
\end{comment}

\end{document}